\documentclass[conference]{IEEEtran}
\usepackage[margin=0.7in]{geometry}

\usepackage[T1]{fontenc}

\usepackage{cite}
\usepackage{graphicx}
\graphicspath{{Figure/}}
\usepackage[cmex10]{amsmath}
\usepackage{amsthm}
\usepackage{amssymb}
\usepackage{algorithmic}
\usepackage{array}
\usepackage{color}
\usepackage{epstopdf}
\usepackage{amsfonts}
\usepackage[ruled,vlined]{algorithm2e}

\usepackage{pgfplots}
\pgfplotsset{compat=1.3}
\usepackage{tikz}							
\usetikzlibrary{shapes}
\usetikzlibrary{spy}
\usetikzlibrary{circuits}
\usetikzlibrary{arrows}
\usepackage{subfig}

\newcommand{\vect}[1]{\boldsymbol{\mathrm{#1}}}
\newcommand{\mat}[1]{\boldsymbol{\mathrm{#1}}}

\newcommand{\tr}{\mathrm{tr}}

\newcommand{\diag}{\text{diag}}

\usepackage{mathtools}

\begin{document}

	\title{Robust Non-Coherent Beamforming for\\ FDD Downlink Massive MIMO}
	
	\author{\IEEEauthorblockN{François Rottenberg\IEEEauthorrefmark{1}, Ming-Chun Lee\IEEEauthorrefmark{1}, Thomas Choi\IEEEauthorrefmark{1}, Jianzhong Zhang\IEEEauthorrefmark{2} and Andreas F. Molisch\IEEEauthorrefmark{1},
		}
		\IEEEauthorblockA{\IEEEauthorrefmark{1}University of Southern California, Los Angeles, CA, USA,
		}
		\IEEEauthorblockA{\IEEEauthorrefmark{2}Samsung Research America, Richardson, TX, USA
		}
	}
	
	\maketitle

	\begin{abstract}
		Designing beamforming techniques for the downlink (DL) of frequency division duplex (FDD) massive MIMO is known to be a challenging problem due to the difficulty of obtaining channel state information (CSI). Indeed, since the uplink-downlink bands are disjoint, the system cannot rely on channel reciprocity to estimate the channel from uplink (UL) pilots as in time division duplexing (TDD) system. Still, in this paper, we propose original designs for robust beamformers that do not require any feedback from the users and only rely on the transmission of UL pilots. The price to pay is that the beamformer is \textit{non-coherent} in the sense that it does not leverage full knowledge of the phase of each multipath component. A large variety of novel designs are proposed under different criterion and partial phase knowledge.
	\end{abstract}
	
	\begin{IEEEkeywords}
		Massive MIMO, FDD, Robust Beamforming, Non-Coherent.\vspace{-1.5em}
	\end{IEEEkeywords}
		\linespread{0.85}
	\section{Introduction}\label{section:Introduction}

	The large benefits promised by massive MIMO technology critically rely on the knowledge of CSI at the base station (BS). In FDD, the BS cannot rely on channel reciprocity from UL pilots to obtain this CSI since the UL and DL bands are disjoint. This makes the acquisition of DL CSI much more challenging. Many works have been proposed in the literature to address this problem. Most of them rely on a limited amount of feedback from the users based on DL pilots, \textit{e.g.}, \cite{Adhikary2013,Rao2014,Barzegar2019}. However, this has the disadvantage of generating an overhead and requires the channel to be static between transmission of the DL pilots, feedback, and use of the CSI in beamforming. An interesting alternative is channel extrapolation from the UL to the DL band as it completely removes the overhead. Such extrapolation can be realized by using high-resolution techniques \cite{Molisch2011,Yang2018}. The theoretical performance of such methods was studied by us in \cite{rottenberg2019channel} and we showed that the MSE of the extrapolated channel scales with the ratio of the extrapolation range and the uplink bandwidth.
	
	However, high-resolution techniques suffer from a large implementation complexity. Moreover, channel extrapolation becomes increasingly difficult for large FDD separation between UL-DL bands, as the phase of each path becomes inaccurate. In this paper, we investigate the design and the performance of robust non-coherent beamformer at the BS, which is particularly justified when the phase of each path is completely or partially unknown. Though suboptimal (except in particular cases), non-coherent beamformers can be designed relying only on UL pilots (through the spatial covariance of the channel) and thus remove the need of a DL feedback, while leveraging the knowledge of only partial phase knowledge. This problem was actually already studied in \cite{Zetterberg1995,Hugl1999}.
	
	The main novelty of this paper firstly resides in a more general formulation including \textit{partial} phase knowledge through their covariance matrix and secondly in analyzing a large variety of robustness criteria. Indeed, various designs are considered, first in the single-user (SU) case, including maximizing the stationary signal-to-noise ratio (SNR) or the worst-case performance. In the multiple-user (MU) case, both zero forcing (ZF) and signal-to-noise-and-leakage ratio (SNLR) are investigated.

	
	\textbf{Notations}: Vectors and matrices are denoted by bold lowercase and uppercase letters, respectively. Superscripts $^*$, $^T$ and $^H$ stand for conjugate, transpose and Hermitian transpose operators. The symbols $|.|$, $\|.\|^2$, $\jmath$, $\tr(.)$, $\mathbb{E}(.)$, $\Im(.)$ and $\Re(.)$ denote the complex modulus, the Frobenius norm, imaginary unit, trace, expectation, imaginary and real parts, respectively. Vector $\vect{v}_{\mathrm{max}}(\mat{X})$ is the eigenvector of matrix $\mat{X}$ associated to its largest eigenvalue, denoted by $\lambda_{\mathrm{max}}(\mat{X})$. 

	\section{Channel Model}
	\label{section:system_model}

\begin{figure}[!t]  
		\centering
		
		\resizebox{0.35\textwidth}{!}{%
			{\includegraphics[clip, trim=0cm 12.5cm 17.5cm 0cm, scale=1]{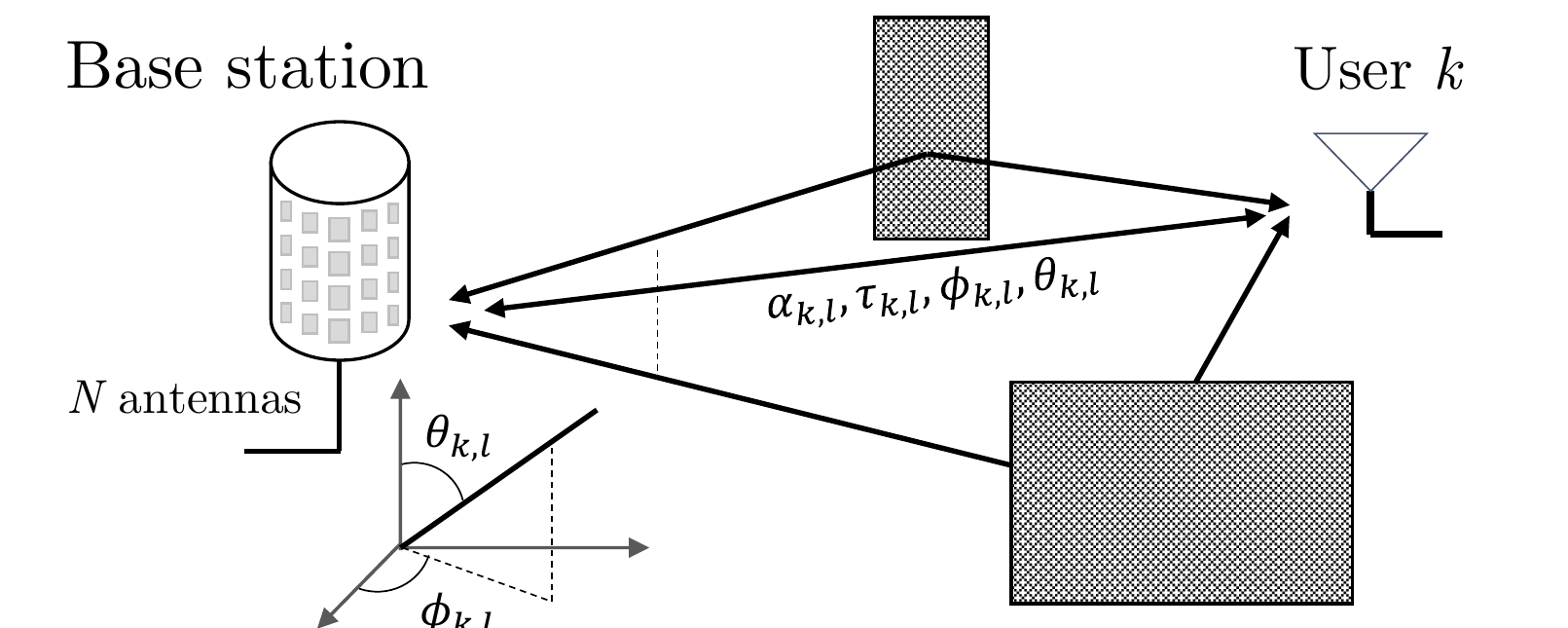}} 
		}
		\caption{Massive MIMO multipath propagation channel.}
		\label{fig:channel_model}
					\vspace{-2.5em}
	\end{figure}

	We consider a massive MIMO system composed of a single BS, equipped with a total of $N$ antennas and $K$ single-antenna users. The BS and the users communicate using the cyclic prefix-orthogonal frequency division multiplexing (CP-OFDM) modulation. We denote by $\vect{h}_k(f,t) \in \mathbb{C}^{N\times 1}$ the channel vector of user $k$, evaluated at subcarrier frequency $f$ and at time $t$. As depicted in Fig.~\ref{fig:channel_model}, we assume that $\vect{h}_k(f,t)$ is composed of a total of $L_k$ multipath components and the $l$-th path is fully characterized by its delay $\tau_{k,l}$, its Doppler shift $\nu_{k,l}$, its complex gain $\alpha_{k,l}$ and its azimuth and elevation angles $\phi_{k,l}$ and $\theta_{k,l}$ at the BS,
	\begin{align}
	\vect{h}_k(f,t)=\sum_{l=1}^{L_k} {\alpha}_{k,l}  \vect{a}({\phi}_{k,l},{\theta}_{k,l},f) e^{-\jmath 2\pi f {\tau}_{k,l}+\jmath 2\pi t \nu_{k,l} }, \label{eq:channel_frequency_response}
	\end{align}
	where $\vect{a}({\phi},{\theta},f)  \in \mathbb{C}^{N\times 1}$ is the array steering vector evaluated in direction $({\phi},{\theta})$ and at frequency $f$. We refer the interested reader to \cite{rottenberg2019performance} for more information on the underlying assumptions of this model.
	In the following, we rewrite (\ref{eq:channel_frequency_response}) as
	\begin{align}
	\vect{h}_k(f,t)=\sum_{l=1}^{L_k} \vect{a}_{k,l}(f,t) e^{\jmath \epsilon_{k,l}(f,t)}, \label{eq:non_coherent_channel}
	\end{align}
	where $\vect{a}_{k,l}(f,t)$ is the spatial signatures of path $l$ and $\epsilon_{k,l}(f,t)$ is the so-called \textit{instantaneous phase uncertainty} of path $l$. We refer to the previous model as \textit{non-coherent} channel model. This is motivated by the fact that the phases $\epsilon_{k,l}(f)$ of each path are highly sensitive variables. In the UL band, if pilots are regularly sent by user $k$, the phases can be accurately estimated. However, if the BS uses a different band for the DL, as in FDD operation, the estimation/extrapolation of the phase $\epsilon_{k,l}(f)$ is much more intricate \cite{rottenberg2019performance}. Note that, in the special case $\epsilon_{k,l}(f,t)=0$, no uncertainty remains and the model becomes coherent again.
	
	On the other hand, the spatial signatures $\vect{a}_{k,l}(f,t)$ appearing in (\ref{eq:non_coherent_channel}) have a more deterministic nature, as they depend on the power of each path and its incoming direction $({\phi}_{k,l},{\theta}_{k,l})$. Moreover, the vectors $\vect{a}_{k,l}(f,t)$ can 
	be estimated from UL pilot estimates and extrapolated in the DL band, \textit{e.g.}, for FDD operations. This is why we will assume in the remainder of this paper that we have perfect knowledge of the spatial signatures of each user. 
	
	In the following, we consider the transmission of an OFDM symbol at a specific subcarrier frequency and a multicarrier symbol. For clarity and without loss of generality, we drop the frequency-time dependence in (\ref{eq:non_coherent_channel})
	\begin{align*}
		\vect{h}_k=\sum_{l=1}^{L_k} \vect{a}_{k,l} e^{\jmath \epsilon_{k,l}}=\mat{A}_k \vect{v}_k,
	\end{align*}
	where $\mat{A}_k\triangleq\begin{pmatrix}
	\vect{a}_{k,1},...,\vect{a}_{k,L_k} \end{pmatrix} \in \mathbb{C}^{N\times L_k}$ and $\vect{v}_k\triangleq\begin{pmatrix}
	e^{\jmath \epsilon_{k,1}},..., e^{\jmath \epsilon_{k,L_k}} 
	\end{pmatrix}^T\in \mathbb{C}^{L_k\times 1}$. 
	In DL, the BS uses the beamforming vector $\vect{g}_k \in \mathbb{C}^{N\times 1}$ to transmit the symbol $s_k$ to user $k$. The symbols $s_k$ have variance $p_k$ and are uncorrelated. The beamforming vectors are normalized so that $\|\vect{g}_k\|^2\leq 1$. The received sample at user $k$ can then be written as
	\begin{align}
	r_k=\underbrace{\vect{g}_k^H \vect{h}_k s_k}_{\mathrm{Useful\ Signal}} + \underbrace{\sum_{k'\neq k} \vect{g}_{k'}^H \vect{h}_{k} s_{k'}}_{\mathrm{Inter-User\ Interference}}  +\underbrace{w_k}_{\mathrm{Noise}}, \label{eq:received_signal}
	\end{align}
	where $w_k$ is zero mean additive complex circularly symmetric white Gaussian noise of variance $\sigma^2$. 
	
	\section{Robust Beamforming}
	The challenge that we address in this section is the design of the beamforming vector $\vect{g}_k$ for fixed power $p_k$ given the uncertainty on $\vect{h}_{k}$ through the phases $\epsilon_{k,l}$, that we consider as random variables. In the following, we assume that the users are capable of estimating their equivalent channel $\vect{g}_k^H \vect{h}_k$ during data reception and thus can perform a phase rotation on $r_k$ (coherent detection), which is implicitly captured by the considered objective functions. We first focus on the single-user case before addressing the multi-user case.
	\subsection{Single-User Case}
	In this case, no inter-user interference (IUI) is present and we drop the subscript $k$ in every user-dependent notation for clarity. Equation (\ref{eq:received_signal}) simplifies to
	\begin{align}
	r=\vect{g}^H \vect{h} s+w,\label{eq:channel_SU}
	\end{align}
	where $\vect{h}=\mat{A}\vect{v}$ and $\vect{v}=(e^{\jmath \epsilon_{1}},..., e^{\jmath \epsilon_{L}} 
	)^T$. The single-user problem simplifies to maximizing the beamforming power $|\vect{g}^H \vect{h}|^2$, which provides robustness against additive noise.
	
	\subsubsection{Stationary Beamforming Power Criterion}
\begin{figure}[!t]  
		\centering
		
		\resizebox{0.35\textwidth}{!}{%
			{\includegraphics[clip, trim=0cm 14cm 24cm 0cm, scale=1]{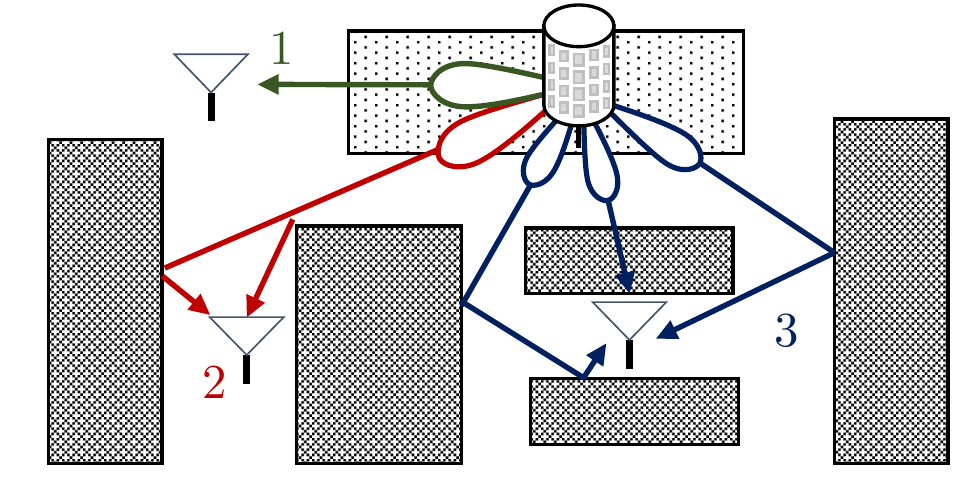}} 
		}

		\caption{Different potential user locations. Location 1 is pure line-of-sight. Location 2 has two paths having equal spatial signatures but different instantaneous phases. Location 3 has three spatially well separated paths with similar power.}
		\label{fig:Single_user_scenario}
					\vspace{-2em}
	\end{figure}

	Since the instantaneous value of $\vect{h}$ is not known we propose first to maximize the so called \textit{stationary} beamforming power obtained by averaging $|\vect{g}^H \vect{h}|^2$ over the statistics of the instantaneous phase uncertainties $\epsilon_{l}$. Given the phase correlation matrix $\mat{R}=\mathbb{E} (\vect{v}\vect{v}^H)$, we find $ \mathbb{E} (\vect{h}\vect{h}^H)=\mat{A}\mat{R}\mat{A}^H$ and the optimization can be formulated as
	\begin{align}
	\vect{g}_{\mathrm{Station.}}&=\arg \max_{\vect{g}}\ \mathbb{E}|\vect{g}^H \vect{h}|^2\quad \text{s.t.}\quad \|\vect{g}\|^2 \leq 1\nonumber \\
	&=\arg \max_{\vect{g}} \vect{g}^H  \mat{A}\mat{R}\mat{A}^H \vect{g}\quad \text{s.t.}\quad \|\vect{g}\|^2\leq 1. \label{eq:max_stationary_beamforming_power}
	\end{align}
	The solution can be easily found as the dominant eigenvector of matrix $\mat{A}\mat{R}\mat{A}^H$
	\begin{align}
	\vect{g}_{\mathrm{Station.}}&=\vect{v}_{\mathrm{max}}(\mat{A}\mat{R}\mat{A}^H), \label{eq:sol_SU_station}
	\end{align}
	which physically implies that the BS should form a beam in the dominant spatial direction, resulting in a \textit{stationary} beamforming power equal to the largest eigenvalue of $\mat{A}\mat{R}\mat{A}^H$ that we denote by $\lambda_{\mathrm{max}}(\mat{A}\mat{R}\mat{A}^H)$. This performance has to be compared to the one of the optimal coherent beamforming vector $\vect{g}=\vect{h}/\|\vect{h}\|$, obtained as a special case of above derivation for an all-one $\mat{R}$ matrix and achieving a \textit{stationary} beamforming power $\mathbb{E}\left(\|\vect{h}\|^2\right)=\tr(\mat{A}\mat{R}\mat{A}^H)$. Denoting by $\tilde{r}$ the rank of $\mat{A}\mat{R}\mat{A}^H$, we can easily find the following upper and lower bounds for $\lambda_{\mathrm{max}}(\mat{A}\mat{R}\mat{A}^H)$
	\begin{align}
	\frac{\tr(\mat{A}\mat{R}\mat{A}^H)}{\tilde{r}}\leq \lambda_{\mathrm{max}}(\mat{A}\mat{R}\mat{A}^H)\leq \tr(\mat{A}\mat{R}\mat{A}^H). \label{eq:inequality_SU}
	\end{align}
	In practice, knowledge of $\mat{R}$ might be difficult to obtain. The most pessimistic case is then to consider i.i.d. uniformly distributed in $[0,2\pi]$ phases, giving $\mat{R}=\mat{I}_L$. Depending on the conditioning of $\mat{A}\mat{A}^H$, the two following extreme cases can be distinguished: i) the upper bound of (\ref{eq:inequality_SU}) is reached by the non-coherent beamformer in the case of a rank-one channel, \textit{i.e.}, $\tilde{r}=1$, which is likely to happen if the user has one strongly dominant spatial direction as user locations 1 and 2 in Fig.~\ref{fig:Single_user_scenario}. In that case, the coherent and non-coherent beamforming gains are equivalent in the stationary/"average" sense. ii) the lower bound of (\ref{eq:inequality_SU}) arises if the channel has $\tilde{r}$ orthogonal equipowered spatial directions. For instance, if the three paths arriving at user location 3 in Fig.~\ref{fig:Single_user_scenario} are spatially well separated and of equivalent power, the non-coherent beamformer will perform about three times ($\approx 4.8$ dB) worse than the coherent one. In the most pessimistic case, the channel has $\tilde{r}=N$ equipowered orthogonal spatial modes and the \textit{stationary} performance corresponds to the one of the uniform beamformer $\vect{g}=\vect{1}/\sqrt{N}$ implying that the knowledge of spatial signatures does not provide any advantage over an agnostic BS.
	
	\subsubsection{Worst-Case Beamforming Power Criterion} \label{subsubsection:worst_case_SU}
	Only the \textit{stationary} beamforming power is maximized in (\ref{eq:max_stationary_beamforming_power}) while no guarantees are provided on the instantaneous one. In this sense, the user might be subject to fades across time and frequency. To avoid this, we could instead design the beamformer vector in order to maximize the \textit{worst-case} performance, which we formalize as
	\begin{align}
	\vect{g}_{\mathrm{Worst}}&=\arg \max_{\vect{g}}\ \min_{\epsilon_1,...,\epsilon_L} |\vect{g}^H \vect{h}|^2\quad \text{s.t.}\ \|\vect{g}\|^2 \leq 1, \label{eq:max_worst_case}\\
	&=\arg \max_{\vect{g}}\ \min_{\epsilon_1,...,\epsilon_L} |\sum_{l=1}^L\vect{g}^H  \vect{a}_l e^{\jmath \epsilon_l}|^2\quad \text{s.t.}\ \|\vect{g}\|^2 \leq 1. \nonumber
	\end{align}
	At first sight, the above problem is not trivial to solve but we can make a few first observations. Depending on the channel environment, a zero \textit{worst-case} beamforming power might be unavoidable. For instance, consider the case of $L=2$ paths coming from the same direction, having the same power but different delays leading to equal spatial signatures $\vect{a}_1=\vect{a}_2$ but different phases $\epsilon_1$ and $\epsilon_2$ (see user location 2 in Fig.~\ref{fig:Single_user_scenario}). The \textit{worst-case} arises if the two paths have opposite phases, \textit{i.e.}, $\epsilon_1 = -\epsilon_2 \mod 2\pi$ and is independent of the beamforming vector $\vect{g}$. If such a case occurs in practice, the BS can either (i) still use a non-coherent beamformer and rely on coding across time and frequency to recover from potential fades; (ii) transmit one pilot in the direction $\vect{a}_1$, obtain some feedback information from the user and perform coherent beamforming. 
	
	We can also see that, if at least one path is orthogonal to all others (as user location 3 in Fig.~\ref{fig:Single_user_scenario}), it is possible to avoid a zero \textit{worst-case} by using a beamforming vector being the matched filter of the orthogonal path spatial signature. These observations illustrate that a specific propagation scenario might be optimistic regarding the \textit{stationary} beamforming power while performing very bad in the \textit{worst-case} sense and vice-versa, as illustrated by user locations 2 and 3 in Fig.~\ref{fig:Single_user_scenario}. Note also that user location 1 performs very well regarding both criteria. Furthermore, investigating further the problem (\ref{eq:max_worst_case}) and assuming that $\epsilon_l \in [0,2\pi],\forall l$, it is possible to show that it is equivalent to solving 
	\begin{align*}
	\vect{g}_{\mathrm{Worst}}&=\arg \max_{l',\vect{g}}\  |\vect{g}^H  \vect{a}_{l'}|-  \sum_{l=1, l\neq l'}^L|\vect{g}^H  \vect{a}_l|\quad \text{s.t.}\ \|\vect{g}\|^2 \leq 1, \nonumber
	\end{align*}
	which intuitively implies that the \textit{worst-case} arises if the amplitude related to the strongest path is affected by destructive interference from all other paths. If the value of the objective function is zero, it implies that a zero \textit{worst-case} is unavoidable. One can check that, for a fixed $l'$, the objective has the form of a difference of convex (DC) functions in $\vect{g}$, which is non convex but can be minimized using numerical methods \cite{Yuille2003}. Since $L$ is finite and known, one can solve the problem for each $l'\in \{1,...,L\}$ and find then optimal $l'$. This is summarized in the Algorithm~1.
	\begin{algorithm} 
		For each $l'\in \{1,...,L\}$, iteratively solve

		$\min_{\vect{g}}\ -|\vect{g}^H  \vect{a}_{l'}|+  \sum_{l=1, l\neq l'}^L|\vect{g}^H  \vect{a}_l|\quad\text{s.t.}\ \|\vect{g}\|^2 \leq 1 $
		
		(0) Initialize $\vect{g}^{(0)}$ and $n=0$.\\
		
		(1) Linearize $-  |\vect{g}^H  \vect{a}_{l'}|$ around $\vect{g}=\vect{g}^{(n)}$ by using subgradient and update $\vect{g}^{(n+1)}$ by solving the convexified problem.
		
		(2) $n\rightarrow n+1$, go back to step (1) until convergence.\\	
		
		 \textbf{Solution}: Keep optimal $\vect{g}$ over $l'$.
		\caption{Max worst-case beamforming power}
					
	\end{algorithm}

	\subsection{Multi-User Case}
	
	In the multi-user case, according to (\ref{eq:received_signal}), the user $k$ might be affected from IUI which should be properly taken into account. We first consider the zero forcing design ensuring that the interference is completely canceled. Secondly, we discuss more general designs.
	
	\subsubsection{Zero Forcing Design}
	Looking at (\ref{eq:received_signal}), a straightforward choice to completely remove the IUI is to ensure that $\vect{g}_k^H \vect{h}_{k'}=0,\forall (k,k')$ s.t. $k\neq k'$, implying that the signal intended to each user does not interfere with other users. Since $\vect{h}_{k'}=\mat{A}_{k'}\vect{v}_{k'}$, this can be satisfied if $\vect{g}_k^H$ lies in the left null space of the intereference matrix $\mathring{\mat{A}}_{k}\triangleq(\mat{A}_1,...,\mat{A}_{k-1},\mat{A}_{k+1},...,\mat{A}_K)$ of dimension $N\times \mathring{L}_k$ with $\mathring{L}_k\triangleq\sum_{k'=1,k'\neq k}^KL_{k'}$. Let us denote the rank of $\mathring{\mat{A}}_{k}$ by $\mathring{r}_k$. We define its singular value decomposition (SVD) as
	\begin{align*}
	\mathring{\mat{A}}_{k}&=\begin{pmatrix}
	\mathring{\mat{U}}_{k,1}& \mathring{\mat{U}}_{k,2}
	\end{pmatrix} \begin{pmatrix}
	\mathring{\mat{\Sigma}}_k & \mat{0}_{\mathring{r}_k\times \mathring{L}_k- \mathring{r}_k}\\
	\mat{0}_{N-\mathring{r}_k\times \mathring{r}_k} & \vect{0}_{N-\mathring{r}_k\times \mathring{L}_k- \mathring{r}_k}
	\end{pmatrix} 	\mathring{\mat{V}}_{k}^H,
	\end{align*}
	where $\mathring{\mat{U}}_{k,1}\in \mathbb{C}^{N\times \mathring{r}_k}$, $\mathring{\mat{U}}_{k,2}\in \mathbb{C}^{N\times N-\mathring{r}_k}$, $\mathring{\mat{\Sigma}}_k\in \mathbb{R}^{\mathring{r}_k \times \mathring{r}_k}$ and $\mathring{\mat{V}}_k\in \mathbb{C}^{\mathring{L}_k \times \mathring{L}_k}$. The ZF condition implies that $\vect{g}_k$ should adopt the following generic form
	\begin{align*}
	\vect{g}_k&=\mat{P}_k\tilde{\vect{g}}_k,\quad \mat{P}_k\triangleq \mathring{\mat{U}}_{k,2} \mathring{\mat{U}}_{k,2}^H \in \mathbb{C}^{N\times N}.
	\end{align*}
	We refer to the projection matrix $\mat{P}_k$ as a \textit{pre-beamforming} matrix (similarly as was done in \cite{JSDM} for separating groups of users). Note that a projection matrix satisfies $\mat{P}_k^2=\mat{P}_k$ and $\mat{P}_k^H=\mat{P}_k$. If the ZF criterion is fulfilled, the received signal at user $k$, free from IUI, is given by
	\begin{align}
	r_k=\tilde{\vect{g}}_k^H\mat{P}_k \vect{h}_k s_k+w_k=\tilde{\vect{g}}_k^H \tilde{\vect{h}}_k s_k+w_k, \label{eq:ZF_equ_channel}
	\end{align}
	where $\tilde{\vect{h}}_k\triangleq\mat{P}_k \vect{h}_k=\mat{P}_k \mat{A}_{k} \vect{v}_{k}$ is the equivalent channel after \textit{pre-beamforming}. Note that (\ref{eq:ZF_equ_channel}) is freed from IUI and has a similar shape as (\ref{eq:channel_SU}) in the single-user case with an equivalent channel $\tilde{\vect{h}}_k$ instead of $\vect{h}_k$. This implies that the techniques developed in the single-user case can be straightforwardly applied in the multi-user case, \textit{i.e.}, the beamforming vector is then obtained by: (i) replacing the expression of matrix $\mat{A}_k$ and vector $\vect{a}_{k,l}$ in the single-user case by matrix $\mat{P}_k \mat{A}_{k}$ and vector $\mat{P}_k \vect{a}_{k,l}$ respectively, (ii) finding the optimal $\tilde{\vect{g}}_{k}$, and (iii) applying the \textit{pre-beamforming} matrix $\mat{P}_k$. 
	
	For instance, the beamforming vector that would maximize the stationary beamforming power is given by
	\begin{align}
	\vect{g}_{k,\mathrm{Station.}}^{\mathrm{ZF}}&=\mat{P}_k \vect{v}_{\mathrm{max}}(\mat{P}_k\mat{A}_k\mat{R}_k\mat{A}^H_k\mat{P}_k)\nonumber\\
	&=\vect{v}_{\mathrm{max}}(\mat{P}_k\mat{A}_k\mat{R}_k\mat{A}^H_k), \label{eq:ZF_station}
	\end{align}
	where $\mat{R}_k\triangleq \mathbb{E}(\vect{v}_k\vect{v}_k^H) $.
	\begin{figure}[!t]  
		\centering
		
		\resizebox{0.35\textwidth}{!}{%
			{\includegraphics[clip, trim=0cm 14.5cm 24cm 0cm, scale=1]{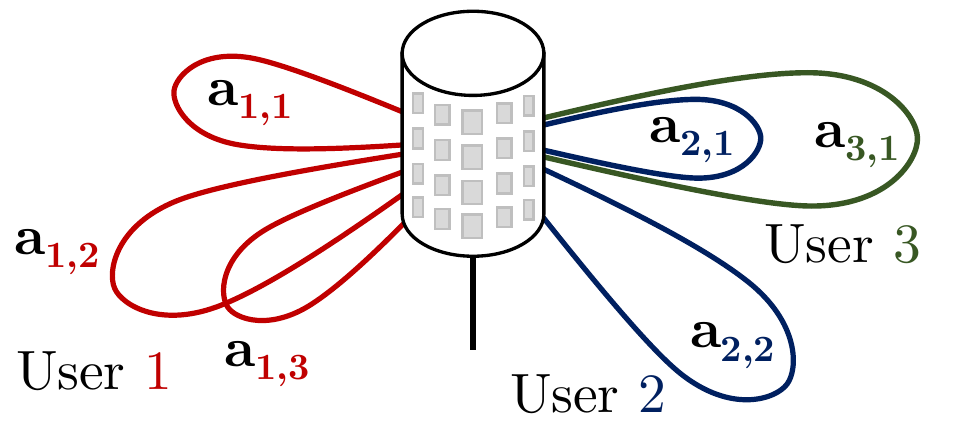}} 
		}
		\caption{Multi-user scenario with $K=3$ users. Vector $\vect{a}_{k,l}$ is the $l$-th spatial signature of the $k$-th user. }
		\label{fig:Multi_user_scenario}
					\vspace{-2em}
	\end{figure}	
	One can wonder about the impact of the \textit{pre-beamforming} matrix $\mat{P}_k$. As depicted in Fig.~\ref{fig:Multi_user_scenario}, depending on channel conditions, three cases can be distinguished: (i) the spatial signatures of user $k=1$ are orthogonal to the ones of all other users, \textit{i.e.}, $\mat{A}_1^H \mathring{\mat{A}}_{1}=\mat{0}$ implying that $\mat{P}_k\mat{A}_k=\mat{A}_k$ and hence, no performance loss is induced, (ii) the spatial signatures of user $k=2$ are not orthogonal to the ones of all other users but cannot be written as an exact linear combination of them, inducing \textit{some} performance loss, and (iii) the spatial signatures of user 3 can be expressed as a linear combination of the ones of of user 2 implying that $\mat{P}_3\mat{A}_3=\mat{0}$ and hence, no degrees of freedom are left after the \textit{pre-beamforming} operation.
	
	The discussion of the previous paragraph induces that a clever multiplexing of the user in the time-frequency-spatial plane should be performed similarly as in the coherent case. If the channels of two users experience closely related spatial structure, it might be preferable to "orthogonalize" them using different time-frequency resources rather than in the spatial domain. Another alternative is the transmission of a few pilots to obtain feedback from the users regarding their common spatial signatures and then perform coherent beamforming \cite{JSDM}. 
	
	
	\subsubsection{Regularized Zero Forcing} In the following, we alleviate the ZF constraint $\vect{g}_k^H \vect{h}_{k'}=\vect{g}_k^H\mat{A}_{k'} \vect{v}_{k'}=0,\forall k\neq k'$ implying that the IUI in (\ref{eq:received_signal}) does not cancel. One should note that relaxing the ZF constraint is particularly useful in the non-coherent case. Indeed, the total number of linear constraints $\tilde{r}_k$ scales not only with the number of users $K$ as in the coherent case but also the number of spatial signatures per-user $L_k$, which can be very large, especially in the case of diffuse multipath components. However, some of these paths might have very small power and hence lead to little interference. Imposing a ZF criterion that completely cancel these paths might be too restrictive while a relaxed criterion can greatly improve the performance.
	
	To relax the ZF criterion, we use the \textit{non-coherent} equivalent of the conventional regularized ZF beamformer \cite{Bjornson2014}. 
	The optimization problem can be written as the maximization of the \textit{stationary} signal-to-leakage-and-noise ratio (SLNR), defined as the ratio of the \textit{stationary} power of the useful signal in (\ref{eq:received_signal}) to the \textit{stationary} power of interference caused to the other users plus noise
	\begin{align*}
	\vect{g}^{\mathrm{RZF}}_{k,\mathrm{Station.}}&=\arg \max_{\vect{g}_k, \|\vect{g}_k\|^2\leq 1}\  \frac{p_k \mathbb{E}|\vect{g}_k^H \vect{h}_{k}|^2}{p_k\sum_{k'\neq k}\mathbb{E}|\vect{g}_k^H \vect{h}_{k'}|^2+\sigma^2}\\
	&=\arg \max_{\vect{g}_k, \|\vect{g}_k\|^2= 1}\  \frac{\vect{g}_k^H \mat{A}_k \mat{R}_k\mat{A}_k^H\vect{g}_k}{\vect{g}_k^H \left( \mathring{\mat{A}}_{k}\mathring{\mat{R}}_{k}\mathring{\mat{A}}_{k}^H+\rho_k\mat{I}_N\right) \vect{g}_k},
	\end{align*}
	where $\rho_k\triangleq \sigma^2/p_k$ is the inverse of the SNR and $\mathring{\mat{R}}_{k}\triangleq \diag({\mat{R}}_{1},...,{\mat{R}}_{k-1},{\mat{R}}_{k+1},...,{\mat{R}}_{K})$ is block diagonal. Since the problem has the form of a generalized Rayleigh quotient, the solution is given by the dominant generalized eigenvector of the matrix pair $(\mat{A}_k\mat{R}_k\mat{A}_k^H,\mathring{\mat{A}}_{k}\mathring{\mat{R}}_{k}\mathring{\mat{A}}_{k}^H+\rho_k\mat{I}_N)$. Since matrix $(\mathring{\mat{A}}_{k}\mathring{\mat{R}}_{k}\mathring{\mat{A}}_{k}^H+\rho_k\mat{I}_N)$ is of full rank, the solution simplifies to the dominant eigenvalue problem
	\begin{align*}
	\vect{g}^{\mathrm{RZF}}_{k,\mathrm{Station.}}&=\vect{v}_{\mathrm{max}} \left( \left( \mathring{\mat{A}}_{k}\mathring{\mat{R}}_{k}\mathring{\mat{A}}_{k}^H+\rho_k\mat{I}_N\right)^{-1}  \mat{A}_k{\mat{R}}_{k}\mat{A}_k^H   \right).
	\end{align*}
	At low SNR, the beamforming vector converges to the single-user solution previously derived in (\ref{eq:sol_SU_station})
	\begin{align*}
	\lim_{\rho_k \rightarrow +\infty}\ \vect{g}^{\mathrm{RZF}}_{k,\mathrm{Station.}}=\vect{v}_{\mathrm{max}} (   \mat{A}_k{\mat{R}}_{k}\mat{A}_k^H   ),
	\end{align*}
	which makes sense as the system is noise limited and IUI is negligible. We now analyze the high SNR regime. To simplify the analysis, we consider the i.i.d. phase case $\mat{R}_{k}=\mat{I}_{L_k},\ \forall k$. Using the SVD of $\mathring{\mat{A}}_{k}$ and the Woodbury matrix inversion lemma, we can rewrite the inverse as
	\begin{align*}
	&( \mathring{\mat{A}}_{k}\mathring{\mat{A}}_{k}^H+\rho_k\mat{I}_N)^{-1}=( \mathring{\mat{U}}_{k,1}\mathring{\mat{\Sigma}}_k^2\mathring{\mat{U}}_{k,1}^H+\rho_k\mat{I}_N)^{-1}\\
	&=\rho_k^{-1} \left(\mat{I}_N - \mathring{\mat{U}}_{k,1} \left(\rho_k \mathring{\mat{\Sigma}}_k^{-2}+ \mat{I}_{\mathring{r}_k}\right)^{-1} \mathring{\mat{U}}_{k,1}^H \right).
	\end{align*}
	Using the fact that $\mat{P}_k=\mat{I}_N - \mathring{\mat{U}}_{k,1} \mathring{\mat{U}}_{k,1}^H$ and taking the limit, we find
	\begin{align*}
	\lim_{\rho_k \rightarrow 0}\ \vect{g}^{\mathrm{RZF}}_{k,\mathrm{Station.}}&=\vect{v}_{\mathrm{max}} \left( \mat{P}_k  \mat{A}_k\mat{A}_k^H   \right).
	\end{align*}
	This result corresponds to the ZF solution obtained in (\ref{eq:ZF_station}) for $\mat{R}_{k}=\mat{I}_{L_k}$, which again makes sense as the system is limited by IUI. 
	
	
	\begin{figure*}[t]
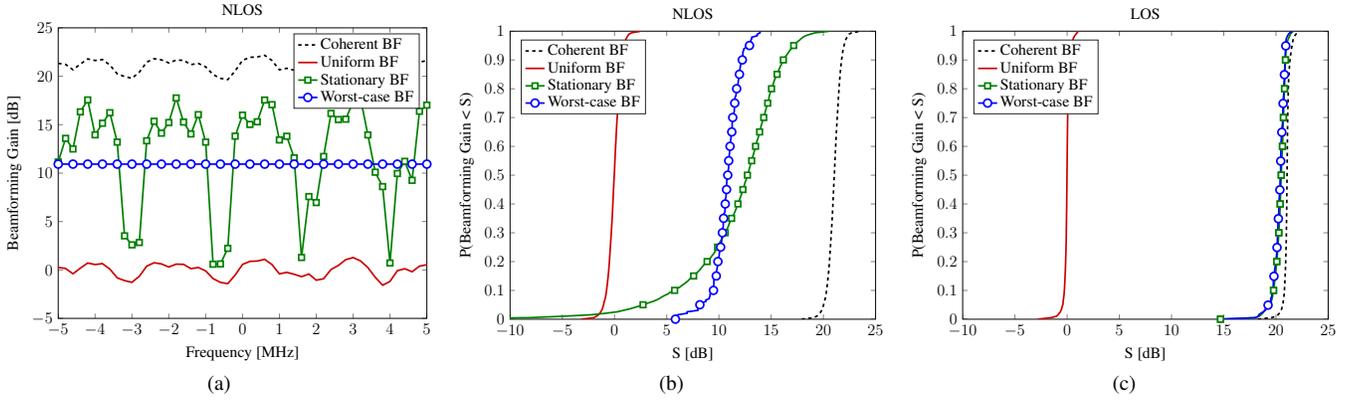

		\centering 
		\subfloat[ ]{
			\resizebox{0.32\textwidth}{!}{\Large 
%
%
\begin{tikzpicture}

\begin{axis}[%
width=4.520833in,
height=3.565625in,
at={(0.758333in,0.48125in)},
scale only axis,
xmin=-5,
xmax=5,
xlabel={Frequency [MHz]},
ymin=-5,
ymax=25,
ylabel={Beamforming Gain [dB]},
title={NLOS},
legend style={legend cell align=left,align=left,draw=white!15!black}
]
\addplot [color=black,dashed,line width=1.5pt]
  table[row sep=crcr]{%
-5	21.335898262257\\
-4.8	21.2250474604274\\
-4.6	20.6746561058837\\
-4.4	21.2568500001581\\
-4.2	21.7887962527264\\
-4	21.6284034540778\\
-3.8	21.7241508677338\\
-3.6	21.1666057410875\\
-3.4	20.2573919286673\\
-3.2	19.9723223320592\\
-3	19.7933794060941\\
-2.8	20.4207280609001\\
-2.6	21.4589624869982\\
-2.4	21.8153289287549\\
-2.2	21.6926745495939\\
-2	21.3759279941907\\
-1.8	21.6719535081764\\
-1.6	21.6475000830983\\
-1.4	21.2045456186423\\
-1.2	21.3266244742723\\
-1	20.9735418139921\\
-0.8	20.1793415410375\\
-0.6	19.7884057172795\\
-0.4	19.6558960132514\\
-0.2	20.5475194069273\\
0	21.6437679109459\\
0.2	21.9579136009843\\
0.4	21.9938591756648\\
0.6	22.1663670927901\\
0.8	21.6254024276971\\
1	20.6669209272237\\
1.2	20.824621464042\\
1.4	20.6248167164297\\
1.6	20.3781641826826\\
1.8	20.6522379152867\\
2	20.0152377580059\\
2.2	20.1643927655601\\
2.4	21.1190047420814\\
2.6	21.5367160655358\\
2.8	22.1356894032382\\
3	22.3485588666129\\
3.2	22.0083641426979\\
3.4	21.3190634240189\\
3.6	20.3194846990574\\
3.8	19.492124369613\\
4	19.8909575582843\\
4.2	20.993435043317\\
4.4	21.1997772172393\\
4.6	20.8851869129323\\
4.8	21.4632043002594\\
5	21.6119817972975\\
};
\addlegendentry{Coherent BF};

\addplot [color=black!20!red,solid,line width=1.5pt]
  table[row sep=crcr]{%
-5	0.263798565778303\\
-4.8	0.1529477639487\\
-4.6	-0.397443590594964\\
-4.4	0.184750303679399\\
-4.2	0.716696556247747\\
-4	0.556303757599145\\
-3.8	0.652051171255154\\
-3.6	0.0945060446088575\\
-3.4	-0.814707767811431\\
-3.2	-1.09977736441948\\
-3	-1.27872029038456\\
-2.8	-0.651371635578617\\
-2.6	0.386862790519554\\
-2.4	0.74322923227622\\
-2.2	0.620574853115207\\
-2	0.30382829771199\\
-1.8	0.599853811697735\\
-1.6	0.575400386619594\\
-1.4	0.132445922163651\\
-1.2	0.254524777793665\\
-1	-0.0985578824865807\\
-0.8	-0.892758155441223\\
-0.6	-1.28369397919921\\
-0.4	-1.41620368322725\\
-0.2	-0.52458028955142\\
0	0.571668214467167\\
0.2	0.885813904505589\\
0.4	0.921759479186073\\
0.6	1.09426739631146\\
0.8	0.553302731218385\\
1	-0.405178769254975\\
1.2	-0.247478232436694\\
1.4	-0.447282980048974\\
1.6	-0.693935513796109\\
1.8	-0.419861781191999\\
2	-1.05686193847279\\
2.2	-0.907706930918583\\
2.4	0.0469050456027239\\
2.6	0.46461636905709\\
2.8	1.0635897067595\\
3	1.27645917013421\\
3.2	0.936264446219233\\
3.4	0.246963727540186\\
3.6	-0.752614997421265\\
3.8	-1.57997532686564\\
4	-1.18114213819434\\
4.2	-0.0786646531616535\\
4.4	0.127677520760572\\
4.6	-0.186912783546369\\
4.8	0.391104603780761\\
5	0.539882100818797\\
};
\addlegendentry{Uniform BF};

\addplot [color=black!50!green,solid,line width=1.5pt,mark size=2.5pt,mark=square*,mark options={solid,fill=white}]
  table[row sep=crcr]{%
-5	11.149449895274\\
-4.8	13.5850872629462\\
-4.6	12.4927006183556\\
-4.4	16.3266954103505\\
-4.2	17.5568526127626\\
-4	13.951804236988\\
-3.8	15.1514225507548\\
-3.6	16.2380883930244\\
-3.4	13.2033547370888\\
-3.2	3.52011151921413\\
-3	2.5924800237544\\
-2.8	2.834622500618\\
-2.6	13.3391183257553\\
-2.4	15.3557947174177\\
-2.2	14.1249255483468\\
-2	15.222058518923\\
-1.8	17.7606024522049\\
-1.6	15.2697352447804\\
-1.4	14.0419840809619\\
-1.2	16.036404005479\\
-1	13.2012288240307\\
-0.8	0.591449715252781\\
-0.6	0.621916035802513\\
-0.4	2.22977845660392\\
-0.2	13.8166323179344\\
0	15.9842715484864\\
0.2	15.0246464496509\\
0.4	15.2998655561043\\
0.6	17.5461257334917\\
0.8	17.0682192314364\\
1	13.4248542264822\\
1.2	13.8064393452949\\
1.4	11.5718387818436\\
1.6	1.28856331703786\\
1.8	7.57794266935197\\
2	6.95703592463617\\
2.2	11.7178447308052\\
2.4	16.1614750624398\\
2.6	15.531756506703\\
2.8	15.5707142176719\\
3	17.3773947875173\\
3.2	17.7481684301826\\
3.4	13.9446460158888\\
3.6	10.0932415165624\\
3.8	8.60578711153574\\
4	0.719802357317387\\
4.2	9.95983270682535\\
4.4	11.2152157075677\\
4.6	9.25683486199199\\
4.8	16.3859690906905\\
5	17.0212628893731\\
};
\addlegendentry{Stationary BF};

\addplot [color=blue,solid,line width=1.5pt,mark size=3.5pt,mark=*,mark options={solid,fill=white}]
  table[row sep=crcr]{%
-5	10.9283903450945\\
-4.6	10.9283913311055\\
-4.2	10.9283916369098\\
-3.8	10.9283908873491\\
-3.4	10.9283900021349\\
-3	10.9283900670506\\
-2.6	10.9283910022855\\
-2.2	10.928391661311\\
-1.8	10.9283912342681\\
-1.4	10.9283902478697\\
-1	10.9283899080557\\
-0.6	10.9283906351229\\
-0.2	10.9283915335132\\
0.2	10.9283915033595\\
0.6	10.9283905795759\\
1	10.928389898597\\
1.4	10.9283902934749\\
1.8	10.928391283009\\
2.2	10.9283916487786\\
2.6	10.9283909466525\\
3	10.9283900333047\\
3.4	10.9283900324915\\
3.8	10.9283909428892\\
4.2	10.9283916494529\\
4.6	10.9283912835447\\
5	10.9283902954664\\
};
\addlegendentry{Worst-case BF};

\end{axis}
\end{tikzpicture}%
}
		}
				\subfloat[ ]{
					\resizebox{0.32\textwidth}{!}{\Large \input{SU_CDF_NLOS}}
				}
			\subfloat[ ]{
				\resizebox{0.32\textwidth}{!}{\Large \input{SU_CDF_LOS}}
			}
		\caption{\small Single-user case: (a) Beamforming power across frequency for coherent and non coherent beamformers, for a particular user location in NLOS. (b) Cumulative density function (CDF) over all user locations and frequencies in NLOS. (c) Same as (b) but in LOS condition.}
		\label{fig:SU} 
			\vspace{-2.5em}
	\end{figure*}

	\section{Simulation Results} \label{section:simulation_results}
	
	This section aims at numerically validating the proposed designs. The channel frequency response is generated according to (\ref{eq:channel_frequency_response}). We consider a single OFDM symbol so that the time dependence can be discarded. The bandwidth is set to 10 MHz. The path parameters (gains, delays and angles) are generated by the QuaDRiGa toolbox \cite{Jaeckel2014} according to the 3D-UMa model defined by 3GPP TR 36.873 v12.5.0 specifications \cite{3GPP_TR_36_873v12_5_0}. Both line-of-sight (LOS) and non-line-of-sight (NLOS) types of the model are used. Moreover, we generate the path parameters related to 100 users locations randomly distributed in a radius of 200 meters around the BS; the BS height is 20m above ground. The BS is equipped with a $N=128$ isotropical elements (16 Horiz. $\times$ 8 Vert.) rectangular array. We assume that the spatial signatures $\vect{a}_{k,l}$ are perfectly known at the BS. We consider the pessimistic case ${\mat{R}}_{k}=\mat{I}_{L_k}$.
	
	For computing the worst-case beamforming (BF), we implement Algorithm~1 using the CVX matlab toolbox \cite{cvx}. 
	For the sake of comparison, we also plot the performance of coherent beamforming, \textit{i.e.}, when the phases associated to each spatial signature is perfectly known. In the single-user case, we also include the performance of a uniform beamformer $\vect{g}=\vect{1}/\sqrt{N}$, which represents the case of no channel knowledge.

	
	\subsubsection*{Single-User Case}
	
	We first consider the NLOS case. For one random location, Fig.~\ref{fig:SU}~(a) plots the performance of the proposed non-coherent beamformers. On the one hand, the stationary beamformer performs well on average but is subject to potential fades. On the other hand, the worst-case beamformer has a relatively stable performance across frequency. One can check that there is an approximately 10 dB loss in beamforming power with respect to the coherent beamformer. These observations are further confirmed in Fig.~\ref{fig:SU}~(b) which shows the cumulative density function (CDF) of the beamforming power estimated over all user locations and frequencies. The LOS case is considered in Fig.~\ref{fig:SU}~(c). In that case, the penalty of non-coherent beamformers is much reduced. Indeed, since there is one strongly dominant path, it is sufficient to form a beam in that direction and knowledge of the instantaneous phases of other paths is less important.

	\subsubsection*{Multi-User Case}
	

	\begin{figure}[t]
		\centering 
		\resizebox{0.32\textwidth}{!}{\Large \input{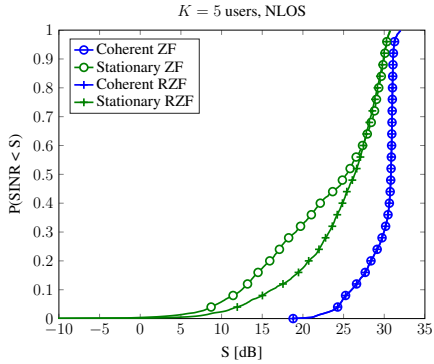}}
		\caption{\small Multi-user case: CDF over 100 user selections in NLOS.}
		\label{fig:MU_NLOS} 
			\vspace{-2em}
	\end{figure}

	\begin{figure}[t]
		\centering 
		\resizebox{0.32\textwidth}{!}{\Large \input{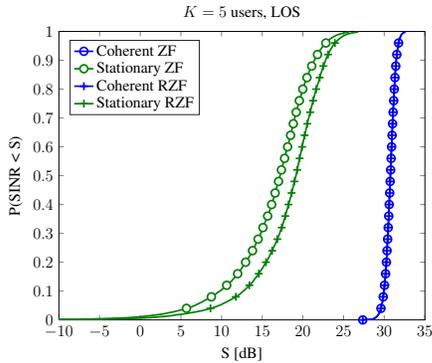}}
		\caption{\small Multi-user case: CDF over 100 user selections in LOS.}
		\label{fig:MU_LOS} 
			\vspace{-2.5em}
	\end{figure}
	
	We now consider the multi-user case with $K=5$ users. From the 100 generated user locations, we select at random 100 combinations of 5 users. The sum of path gains of each user $\alpha_{k,l}$ is normalized to one and the SNR of each user is set to 10 dB, \textit{i.e.}, $\rho_k^{-1}=10$ dB. No power loading and scheduling are implemented even though we expect that they are crucial for non-coherent beamforming. The NLOS and LOS performances are shown in Fig.~\ref{fig:MU_NLOS} and~\ref{fig:MU_LOS} respectively.
		\vspace{-1em}
	
	\section{Conclusion} \label{section:conclusion}
	
	In this paper, we have investigated the design of robust non-coherent beamformers. Most of these designs are in closed-form expressions and easy to compute in terms of complexity and thanks to the fact that they do not rely on any user feedback, directly applicable in FDD massive MIMO systems.
	

	\vspace{-1em}
	
	\footnotesize
	\bibliographystyle{IEEEtran}
	\bibliography{IEEEabrv,IEEEreferences}

\end{document}